# Passivation-Free Ga-Polar AlGaN/GaN Recessed-Gate HEMTs on Sapphire with 2.8 W/mm P$_{OUT}$ and 26.8% PAE at 94 GHz

Ruixin Bai[1], Swarnav Mukhopadhyay[1], Michael Elliott[2], Ryan Gilbert[2], Jiahao Chen[1], Rafael A. Choudhury[1], Kyudong Kim[1], Yu-Chun Wang[1], Ahmad E. Islam[3], Andrew J. Green[3], Shubhra S. Pasayat[1], and Chirag Gupta[1]

[1]Department of Electrical and Computer Engineering, University of Wisconsin-Madison, Madison, WI 53706 USA

[2]KBR, Wright-Patterson AFB, Dayton, OH 45433, USA

[3]Air Force Research Laboratory, Sensors Directorate, Wright-Patterson AFB, Dayton, OH 45433 USA

**Abstract:**

In this work, we demonstrate a passivation-free Ga-polar recessed-gate AlGaN/GaN HEMT on a sapphire substrate for W-band operation, featuring a 5.5 nm Al$_{0.35}$Ga$_{0.65}$N barrier under the gate and a 31 nm Al$_{0.35}$Ga$_{0.65}$N barrier in the gate access regions. The device achieves a drain current density of 1.8 A/mm, a peak transconductance of 750 mS/mm, and low gate leakage with a high on/off ratio of 10$^7$. Small-signal characterization reveals a current-gain cutoff frequency of 127 GHz and a maximum oscillation frequency of 203 GHz. Continuous-wave load-pull measurements at 94 GHz demonstrate an output power density of 2.8 W/mm with 26.8% power-added efficiency (PAE), both of which represent the highest values reported for Ga-polar GaN HEMTs on sapphire substrates and are comparable to state-of-the-art Ga-polar GaN HEMTs on SiC substrates. Considering the low cost of sapphire, the simplicity of the epitaxial design, and the reduced fabrication complexity relative to N-polar devices, this work highlights the potential of recessed-gate Ga-polar AlGaN/GaN HEMTs on sapphire as a promising candidate for next-generation millimeter-wave power applications.

W-band (75–110 GHz) is of increasing importance for high-resolution radar, satellite communications, and emerging wireless systems. GaN HEMTs are particularly well-suited for W-band operation due to their high electron mobility and large breakdown field, which enable both high-power and high-efficiency performance. [1] However, conventional GaN HEMTs with thick passivation layers often exhibit high parasitic capacitances, which impede the simultaneous achievement of high operating frequency and effective dispersion control. To address this issue, recessed-gate structures with thin passivation layer have been explored and successfully demonstrated in N-polar GaN HEMTs, achieving 8.84 W/mm output power (P$_{OUT}$) with 27% power-added efficiency (PAE) on SiC [2], and later 5.8 W/mm P$_{OUT}$ with 38% PAE on sapphire to reduce substrate costs [3]. Despite these impressive results, both the growth and the fabrication of N-Polar GaN HEMTs is more complex than conventional Ga-Polar due to N-Polar orientation's oxygen affinity as well as surface reactivity. [4]-[8]

Ga-polar devices have also been demonstrated at W-band on SiC substrates, including a pre-matched graded-channel HEMT with 2.94 W/mm and 37% PAE [9] and a ScAlN-barrier HEMT with 3.57 W/mm and 24.3% PAE [10], whereas no comparable results have been achieved on sapphire substrate. At the same time, the exploration of recessed-gate structures in Ga-polar AlGaN/GaN HEMTs has been limited, as the growth of a GaN cap typically increases the sheet resistance [11], adversely affecting the RF performance. This limitation was recently addressed by developing a HEMT epitaxial structure with a 31 nm $Al_{0.35}Ga_{0.65}N$ barrier, which enables the formation of gate trenches comparable in depth to N-polar recessed-gate structures without increasing the sheet resistance. [12]

Based on this epitaxial structure, we demonstrate a passivation-free Ga-polar recessed-gate AlGaN/GaN HEMT on sapphire substrate, featuring a 5.5 nm AlGaN barrier beneath the gate. The device exhibits 1.8A/mm current density, 127 GHz cut-off frequency, while maintaining reasonable dispersion control. Under CW load-pull measurements at 94 GHz, it delivers 2.8 W/mm $P_{OUT}$ with 26.8% associated PAE. Considering the lower-cost substrate, the simplicity of the epitaxial structure, and the reduced fabrication complexity of Ga-polar devices, this work highlights the strong potential of Ga-polar HEMTs on sapphire as a competitive platform for future W-band power applications.

The epitaxial layers were grown on Fe-doped semi-insulating GaN buffers on sapphire substrates using metal–organic chemical vapor deposition (MOCVD). [12] As shown in Fig. 1(a), the structure consists of a 1-μm unintentionally doped (UID) GaN channel, a 0.7-nm AlN interlayer and a 31-nm $Al_{0.35}Ga_{0.65}N$ barrier. The fabricated device structure is depicted in **Fig. 1(b)**. Device fabrication began with the n+ ohmic region regrowth process. Following regrowth, mesa isolation was carried out using a $BCl_3/Cl_2$-based reactive ion etching (RIE). A 200 nm $SiO_2$ layer was then deposited by plasma-enhanced chemical vapor deposition (PECVD) to serve as a hard mask. The gate foot was defined by electron-beam lithography (EBL), and the exposed $SiO_2$ in the gate region was etched in an inductively coupled plasma (ICP) system with $CHF_3/CF_4/O_2$ plasma. The AlGaN barrier underneath was subsequently recessed by a low-power $BCl_3/Cl_2$ RIE, leaving a remaining barrier thickness of ~5.5 nm to improve the aspect ratio. After recess formation, a 4 nm $HfO_2$ gate dielectric was deposited by atomic layer deposition (ALD) at 250 °C. The gate foot metal (50 nm TiN) was deposited by ALD at 275 °C. The T-gate head was patterned in a second EBL step and metallized with Cr/Au using electron-beam evaporation. Finally, the $SiO_2$ hard mask was completely removed by buffered oxide etch (BOE), and the ohmic contacts were formed by e-beam evaporation of Ti/Au resulting in a passivation-free device structure. The device under test featured a source-drain spacing ($L_{sd}$) of 0.5 μm, source-gate spacing ($L_{sg}$) of 100 nm and gate length ($L_g$) of 90 nm.

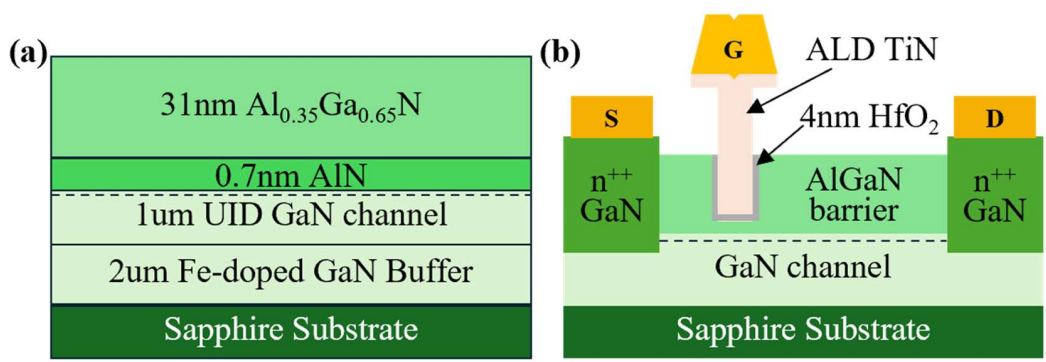

Fig. 1. (a) Epitaxial layer structure and (b)Cross-section schematic view of the passivation-free recessed-gate AlGaN/GaN HEMT.

**Fig. 2** presents the DC output characteristics of the fabricated device. A maximum drain current density of ~1.8 A/mm was achieved, together with a low on-resistance of ~0.41 Ω·mm. These results are enabled by the regrowth-based ohmic contacts, which yielded a small contact resistance of ~0.1 Ω·mm. The high current density further indicates that the deep recess etching, leaving a 5.5 nm AlGaN barrier, did not cause significant channel degradation. The sheet resistance in the recessed region which has been etched was estimated to increase from ~250 Ω/□ to ~320 Ω/□ according to the TLM measurement, which represents only a moderate increase and remains within an acceptable range.

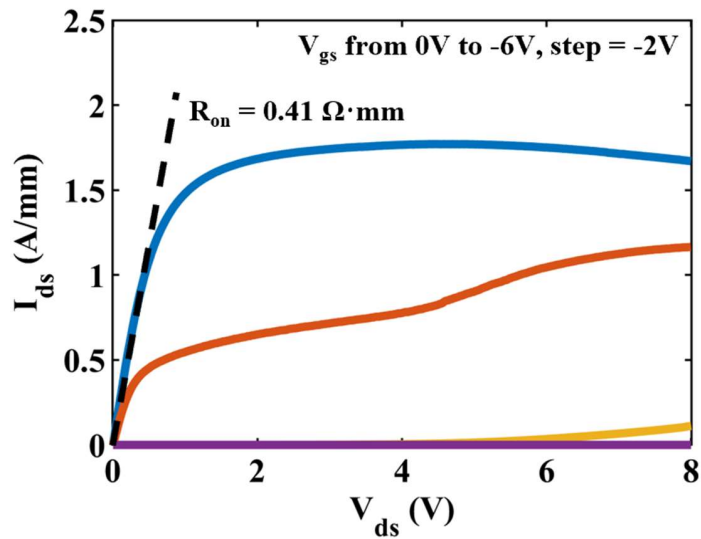

Fig. 2. DC output characteristics of the recessed-gate device.

The transfer I–V characteristics are plotted in both logarithmic and linear scales, as shown in **Fig. 3(a)** and **Fig. 3(b)**. Despite the presence of some short-channel effects, the deep recess structure with thin AlGaN barrier under the gate effectively increases the ratio of the gate length to the total distance between the gate and the channel (aspect ratio), enabling well-defined pinch-off behavior and a high peak transconductance of ~0.75 S/mm. In addition, an on/off current ratio of ~$10^7$ was achieved. This improvement is attributed to the 4 nm $HfO_2$ gate dielectric, which effectively suppresses the leakage current induced by the low work function of TiN, thereby enhancing device stability. [13]

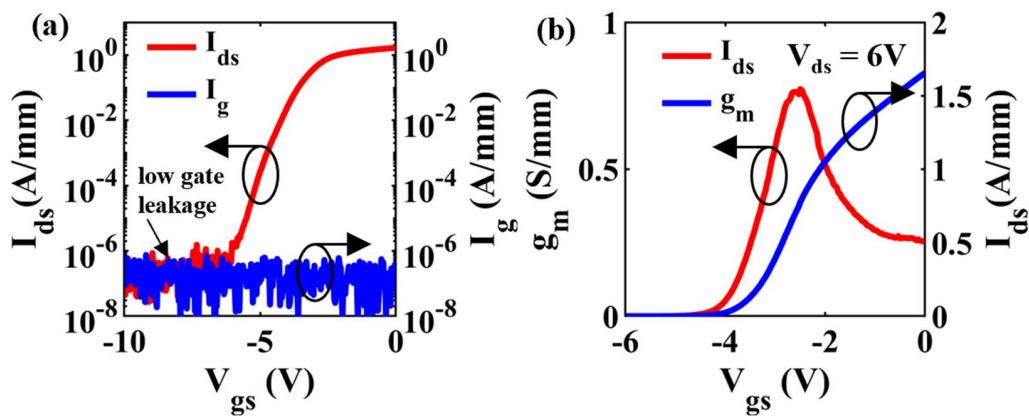

Fig. 3. The transfer characteristics of the recessed-gate device measured at Vds = 6V with (a) linear-scale and (b) log-scale.

Pulsed I–V measurement was performed on a device with $L_{sd}$=700 nm, $L_{sg}$=250 nm, and $L_g$=80 nm. The pulses had a width of 50 μs and a 1% duty cycle, with quiescent voltages of $V_{gsq}$=−3 V and $V_{dsq}$=5 V. As shown in **Fig. 4(a)**, even without any passivation layer, the device demonstrates reasonable dispersion control. Only minor knee walkouts are observed, and no significant current collapse occurs. Slight current degradation (~15%) is present under low drain bias, but it gradually diminishes and becomes negligible as $V_{ds}$ increases. This behavior is primarily attributed to the deep recess structure, which creates a threshold voltage difference between the gate region and the gate access region. This difference partially compensates for the virtual gate effect induced by surface traps in the gate access region, thereby mitigating dispersion. [14], [15]

S-parameters were measured from 100 MHz to 43.5 GHz with SOLT calibration and on-wafer open and short de-embedding. Owing to the absence of any passivation layer, parasitic capacitance associated with additional dielectric layers is avoided, which allows the device to achieve a higher cutoff frequency for a given gate length. As shown in **Fig. 4(b)**, Under quiescent conditions of $V_{gsq}$=−3 V and $V_{dsq}$=6 V, a device with $L_g$=90 nm achieves a cutoff frequency ($f_t$) of 127 GHz and a maximum oscillation frequency ($f_{max}$) of 203 GHz.

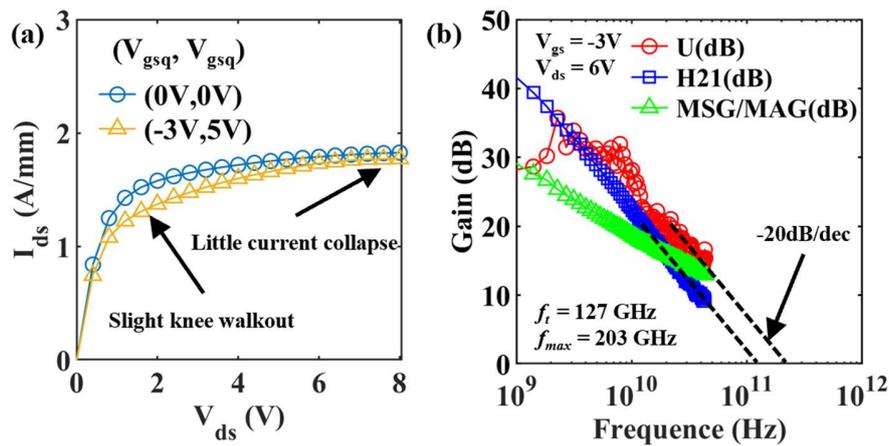

Fig. 4. (a) Pulsed I–V characteristics of the recessed-gate device measured at working condition (b) H21, U and MSG/MAG vs frequency measured at Vgs = -3V and Vds = 6V.

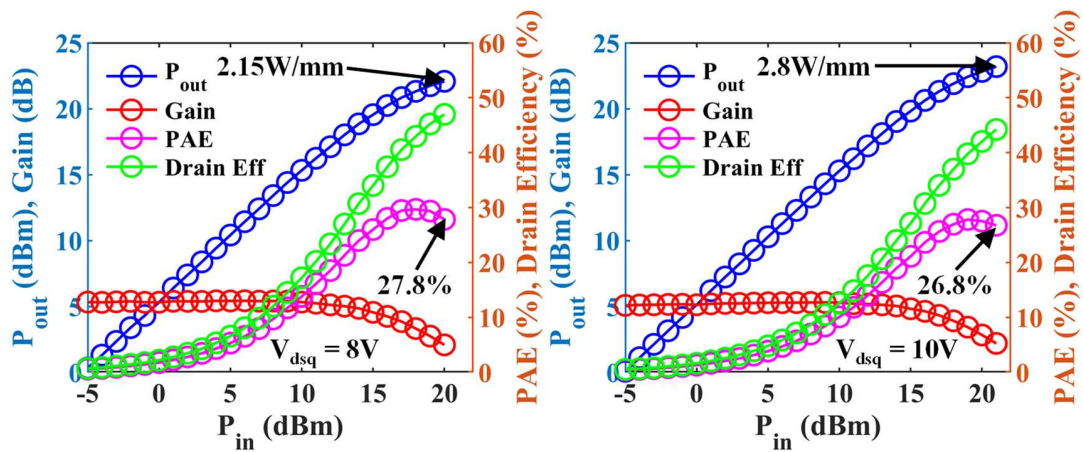

Fig. 5. 94GHz load pull power sweep in CW mode (a) at Vdsq = 8V (b) at Vdsq = 10V.

Continuous-wave (CW) load-pull measurements were conducted at 94 GHz with a passive load-pull system. Prior to the measurements, TRL calibration was performed to ensure accurate extraction of power performance. Since the devices were fabricated on sapphire substrates, the quiescent drain current was maintained at $I_{dsq}$=250 mA/mm to minimize self-heating effects, while the $V_{dsq}$ was increased from 8 V to 10 V. The load reflection coefficient was tuned for maximum PAE. As shown in **Fig. 5**, at $V_{dsq}$=8V, the device achieves an output power density of 2.15 W/mm with a peak power-added efficiency (PAE) of 27.8%. When $V_{dsq}$ is increased to 10 V, the output power density rises to 2.8 W/mm with a PAE of 26.8%. To the best of our knowledge, this is the highest reported W-band performance of Ga-polar GaN channel HEMTs on sapphire, with both $P_{OUT}$ and PAE at 94 GHz exceeding earlier reports by nearly a factor of two.[16]-[18]

**Fig. 6** benchmarks the present results against previously reported Ga-polar [9][10][16]-[24]and N-polar [2][3][25]-[29] GaN HEMTs operating in the 83–95 GHz range. These results compare well with early N-Polar W-band work on sapphire [29] and also some Ga-Polar HEMTs on SiC. With further optimization of fabrication processes, in the future, we anticipate this approach (Ga-Polar HEMT on sapphire) to compare well with other Ga-Polar GaN HEMTs on SiC. Although performance gap remains compared to the most recent N-polar GaN HEMTs on sapphire, the simplicity of our epitaxial structure and the reduced fabrication complexity of Ga-polar devices collectively underscore the strong competitiveness of this work, further highlighting the potential of recessed-gate Ga-polar AlGaN/GaN HEMTs as a promising platform for future W-band power applications.

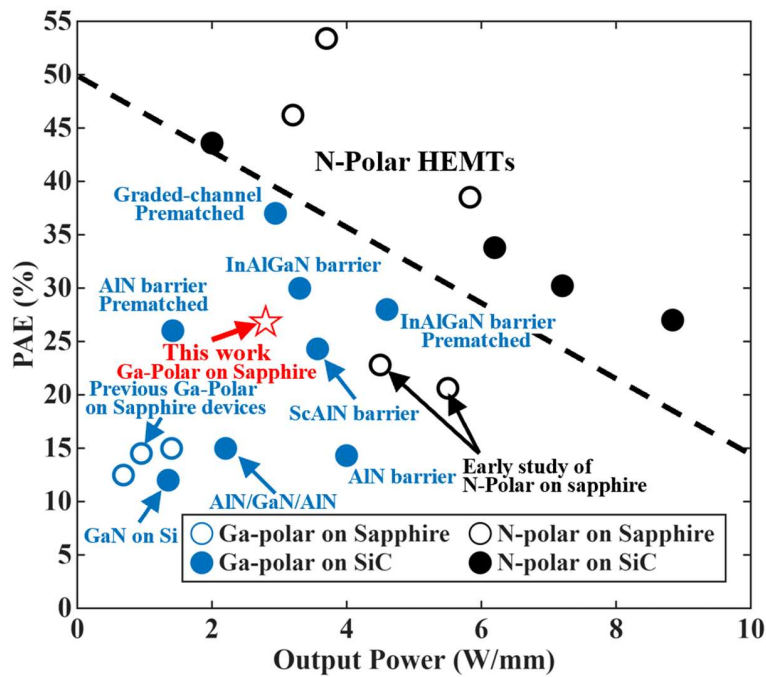

Fig. 6. PAE versus associated $P_{OUT}$ at 83-95 GHz for GaN HEMT. The devices are not pre-matched unless mentioned.

In conclusion, passivation-free Ga-polar recessed-gate AlGaN/GaN HEMTs on sapphire substrates, specifically designed for W-band operation, have been demonstrated in this work. The device exhibits a high peak saturation current density of ~1.8 A/mm and peak transconductance of 0.75 S/mm, along with low gate leakage and a high on/off current ratio of ~$10^7$. The passivation-free recessed-gate structure enables the device to achieve both high-frequency performance (127 GHz $f_t$ and >200GHz $f_{max}$) and reasonable dispersion control (<15% current collapse). CW load-pull measurements at 94GHz further show a $P_{OUT}$ of 2.8 W/mm with an associated PAE of 26.8%, a combination comparable to that of Ga-polar GaN HEMTs on SiC. These results highlight the strong potential of Ga-polar recessed-gate AlGaN/GaN HEMTs on sapphire as a practical and high-performance platform for future W-band power applications.

## Acknowledgment:

This work was supported by AFRL Regional Network - Midwest.